\documentclass{article}
\usepackage{emulateapj,apjfonts}
\input psfig

\newcommand{\etal}{{\it et~al.}}

\newcommand{\Msun}{M_\odot}
\newcommand{\mbh}{$M_\bullet$}

\newcommand{\kms}{$\rm {km}~\rm s^{-1}$}

\newcommand{\vdm}{van~der~Marel}

\begin{document}

\lefthead{BH Mass vs. Velocity Dispersion}
\righthead{Gebhardt~\etal}

\title{A Relationship Between Nuclear Black Hole Mass and Galaxy
Velocity Dispersion}

\author{Karl Gebhardt\altaffilmark{1,2}, Ralf Bender\altaffilmark{3},
Gary Bower\altaffilmark{4}, Alan Dressler\altaffilmark{5},
S.M.~Faber\altaffilmark{2}, Alexei V. Filippenko\altaffilmark{6},
Richard Green\altaffilmark{4}, Carl Grillmair\altaffilmark{7}, Luis C.
Ho\altaffilmark{5}, John Kormendy\altaffilmark{8}, Tod
R. Lauer\altaffilmark{4}, John Magorrian\altaffilmark{9}, Jason
Pinkney\altaffilmark{10}, Douglas Richstone\altaffilmark{10}, and
Scott Tremaine\altaffilmark{11}}

\altaffiltext{1}{Hubble Fellow} 

\altaffiltext{2}{UCO/Lick Observatories, University of California,
Santa Cruz, CA 95064; gebhardt@ucolick.org, faber@ucolick.org}

\altaffiltext{3}{Universit\"ats-Sternwarte, Scheinerstrasse 1,
M\"unchen 81679, Germany; bender@usm.uni-muenchen.de}

\altaffiltext{4}{National Optical Astronomy Observatories, P. O. Box
26732, Tucson, AZ 85726; gbower@noao.edu, green@noao.edu,
lauer@noao.edu}

\altaffiltext{5}{The Observatories of the Carnegie Institution of
Washington, 813 Santa Barbara St., Pasadena, CA 91101;
dressler@ociw.edu, lho@ociw.edu}
 
\altaffiltext{6}{Department of Astronomy, University of California,
Berkeley, CA 94720-3411; alex@astro.berkeley.edu}

\altaffiltext{7}{SIRTF Science Center, 770 South Wilson Ave.,
Pasadena, CA 91125; carl@ipac.caltech.edu}

\altaffiltext{8}{Department of Astronomy, University of Texas, 
RLM 15.308, Austin, Texas 78712; kormendy@astro.as.utexas.edu}
 
\altaffiltext{9}{Institute of Astronomy, Madingley Road, Cambridge
CB3 0HA, England; magog@ast.cam.ac.uk}

\altaffiltext{10}{Dept. of Astronomy, Dennison Bldg., Univ. of
Michigan, Ann Arbor 48109; jpinkney@astro.lsa.umich.edu,
dor@astro.lsa.umich.edu}
 
\altaffiltext{11}{Princeton University Observatory, Peyton Hall,
Princeton, NJ 08544; tremaine@astro.princeton.edu}
 
\begin{abstract}

We describe a correlation between the mass \mbh\ of a galaxy's central
black hole and the luminosity-weighted line-of-sight velocity
dispersion $\sigma_e$ within the half-light radius.  The result is
based on a sample of 26 galaxies, including 13 galaxies with new
determinations of black hole masses from {\it Hubble Space Telescope}
measurements of stellar kinematics. The best-fit correlation is \mbh\
$ = 1.2 (\pm0.2)\times10^8 M_\odot
(\sigma_e/200\hbox{\,\kms})^{3.75\,(\pm0.3)}$ over almost three orders
of magnitude in \mbh; the scatter in \mbh\ at fixed $\sigma_e$ is only
0.30 dex and most of this is due to observational errors. The
\mbh-$\sigma_e$ relation is of interest not only for its strong
predictive power but also because it implies that central black hole
mass is constrained by and closely related to properties of the host
galaxy's bulge.

\end{abstract}
 

\section{Introduction}

Massive black holes at the centers of galaxies are now recognized as a
normal, perhaps ubiquitous, component of elliptical galaxies and
spiral galaxy bulges.  The early evidence is summarized in Kormendy \&
Richstone (1995) and references therein.  Using a heterogeneous set of
36 galaxies, mostly with ground-based spectroscopy and space-based
photometry, Magorrian~\etal\ (1998) argue that {\em all} hot galaxy
components (ellipticals and spiral galaxy bulges) contain central
black holes. Kormendy (1993), Kormendy~\& Richstone (1995), and
Magorrian~\etal\ (1998) also find that black hole mass \mbh\ is
proportional to galaxy mass or luminosity, although with large scatter
(see left panel of Fig.~2). Richstone~\etal\ (1998) outline a
plausible physical framework to discuss the connections between
current black hole mass, galaxy formation, and quasar evolution.  The
set of galaxies with reliable black hole masses is growing rapidly,
mostly through an aggressive campaign of {\it Hubble Space Telescope
(HST)} observations, and this set is now large enough to investigate
the dependence of \mbh\ on galaxy properties, and thus to provide
clues to the role of central black holes in galaxy formation and
evolution.  In this {\it Letter}, we present a new correlation between
line-of-sight velocity dispersion and black hole mass that has very
little intrinsic scatter (probably less than 40\% in \mbh).

\section{The Sample}

Our sample is restricted to galaxies whose black hole masses and
line-of-sight velocity dispersions are well determined.  We use only
two galaxies with maser masses (NGC\,4258 and NGC\,1068), since only
these have well measured dispersions. We also use six galaxies with
black hole masses determined from gas kinematics. Among the galaxies
with black hole masses determined from stellar kinematics, we limit
our sample to those that have three-integral models with {\it HST}
spectroscopy (16 galaxies), plus our Galaxy (Ghez~\etal\ 1998;
Genzel~\etal\ 2000) and M31 (Dressler~\& Richstone 1988, Kormendy
1988). We do {\em not} use 21 galaxies from the sample of
Magorrian~\etal\ (1998) that have only ground-based kinematic data and
two-integral dynamical models. Galaxies with black hole detections
based on isotropic or spherical models (e.g., NGC\,3115:
Kormendy~\etal\ 1996a, NGC\,4594: Kormendy~\etal\ 1996b, NGC\,4486B:
Kormendy~\etal\ 1997) are also excluded until they are re-analyzed
with three-integral models.

The majority of black hole mass estimates in the present sample (14
out of 26) come from galaxies with {\it HST} spectra and preliminary
three-integral axisymmetric dynamical models described by
Gebhardt~\etal\ (2000a, 2000b) and Richstone~\etal\ (2000). Since all
of these galaxies have been observed, analyzed, and modeled with the
same procedures, we expect that the scatter due to systematic errors
will be smaller than in a more heterogeneous sample.

Table~1 presents the data as follows: galaxy name (Column~1), type
(Col.~2), black hole mass and 68\% uncertainty (Col.~3), integrated
line-of-sight velocity dispersion as defined below (Col.~4), distance
in Mpc (Col.~5), and source (Col.~6). The velocity dispersions are
from heterogeneous sources with uncertain errors, but probably most
are accurate to within $\pm$5\%. Most distances come from surface
brightness fluctuations (Tonry~\etal\ 2000); otherwise we use Virgo
infall models to infer the distance (Dekel 2000, private
communication), all scaled to a Hubble constant of
80~\kms\,Mpc$^{-1}$.


\hskip -0.7cm\psfig{file=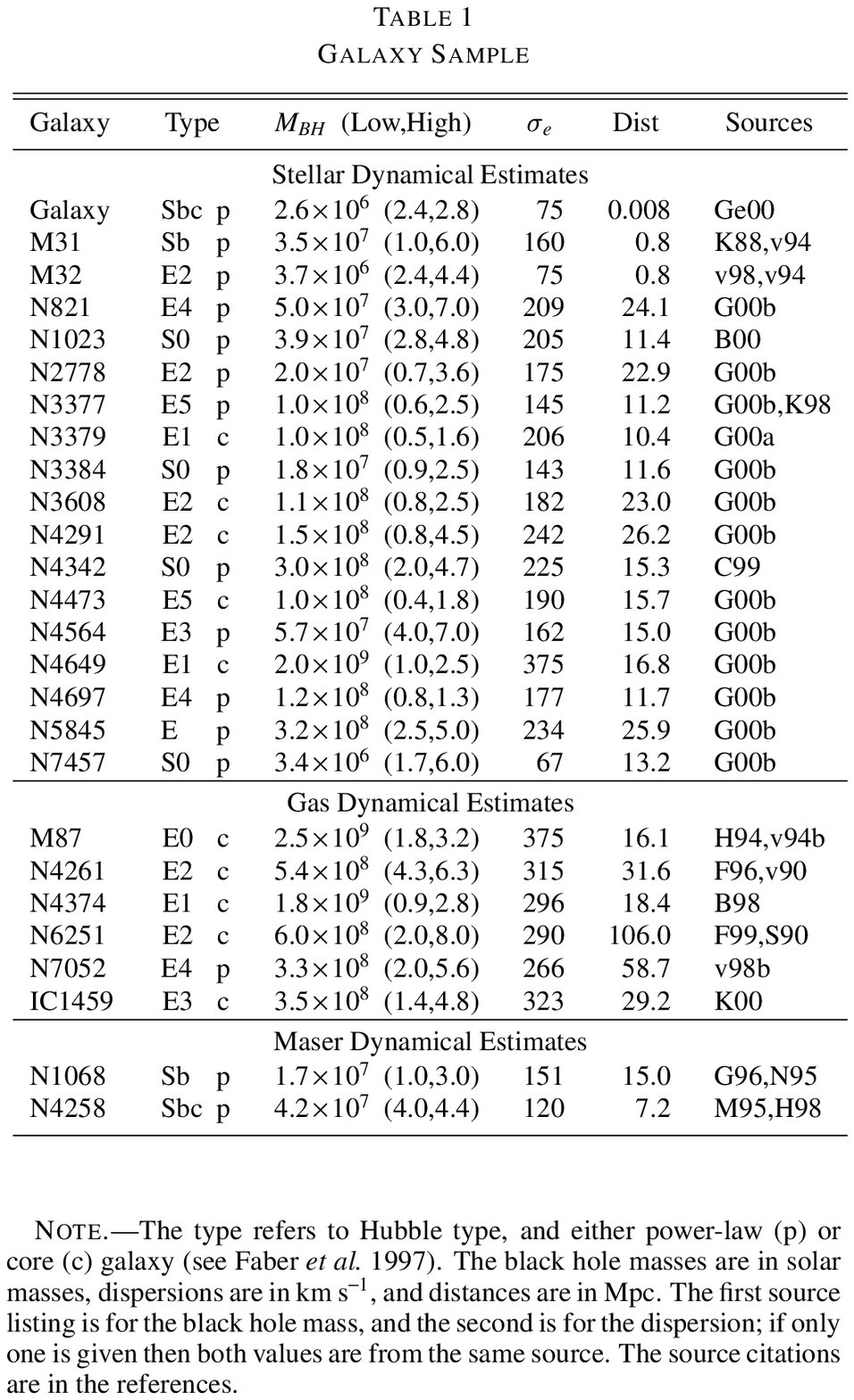,width=16.cm,angle=0}
\vskip -8cm


\vskip -8cm
\section{Estimating the Velocity Dispersion} 

We devote particular care to finding a suitable line-of-sight
dispersion. The traditional estimates of galaxy velocity dispersions
come from the central regions (Faber~\etal\ 1989; J\o rgensen \& Franx
1994), but these are affected by \mbh\ in some galaxies. We focus on
the aperture dispersion (the luminosity-weighted line-of-sight
dispersion inside a radius $R$), since it has a higher signal-to-noise
and is less sensitive to the details of the distribution of orbits
(Richstone \& Tremaine 1984) than is the central dispersion.

For the 13 galaxies studied by Gebhardt~\etal\ (2000b) and
Bower~\etal\ (2000), we have extensive ground-based observations of
the line-of-sight dispersion as a function of radius.  Figure 1
presents the aperture dispersion for these galaxies as a function of
radius, normalized to the effective or half-light radius $R_e$. The
largest variations in the normalized dispersion profile occur in the
center ($R<0.5R_e$) and the outer regions ($R>2R_e$). The variations
among the dispersion profiles in the region $0.5R_e<R<2R_e$ are small,
generally less than 5\%. This result follows because galaxies do not
show dramatic radial variations in dispersion, especially when
integrated over an aperture; the same is generally true of theoretical
models of spherical galaxies. In what follows, we measure
``dispersion'' by the line-of-sight aperture dispersion $\sigma_e$
within $R_e$.


\vskip 0.3cm

\psfig{file=gebhardt.fig1.ps,width=8.5cm,angle=0}
\figcaption[gebhardt.fig1.ps]{Luminosity-weighted line-of-sight
dispersion within an aperture of radius $R$, normalized to its value
at the effective radius $R_e$. Each of the 13 lines represents a
galaxy from Gebhardt~\etal\ (2000b) and Bower~\etal\ (2000).
\label{fig1}}
\vskip 0.3cm


The aperture dispersion includes a contribution from the rotation
(i.e., we measure the rms velocity relative to the systemic velocity,
not relative to the local mean velocity). The measured dispersion
depends on the inclination of the galaxy. The values that we quote
should be regarded as dispersions for galaxies as viewed edge-on,
since several of the galaxies in the sample with the strongest
rotation are nearly edge-on (the Galaxy, M31, NGC~1023, NGC\,3377,
NGC\,3384, and NGC\,4342). For the rest, the inclination is not well
determined; however, the change in dispersion with viewing angle for
ellipticals as judged from models is always $<$ 20\% and on average is
less than 5\% (as calculated from van~der~Marel 1991). 

\section{Results}

Figure~2 plots the black hole mass \mbh\ versus bulge luminosity and
the aperture dispersion $\sigma_e$ at $R_e$. The bulge luminosity
correlation will be discussed in Kormendy~\etal\ (2000b). The
correlation with dipersion is extremely strong, with a correlation
coefficient of 0.93. By a variety of tests, it is much more
significant than the 99\% level. Furthermore, the correlation remains
present in various subsamples---e.g., only galaxies from
Gebhardt~\etal\ (2000b), or only galaxies with gas dynamical or maser
mass estimates. The correlation is robust.

The line in the right panel Figure~2 is a fit to the data assuming
that errors in dispersion measurements are zero and that errors in
$\log$\,\mbh\ are the same for each galaxy.  The best-fit line is
\begin{equation}
M_{\bullet}
=1.2\times10^8~M_\odot\left(\sigma_e\over200\hbox{\,\kms}\right)^{3.75}.
\end{equation}
We have also measured the relation using the seven galaxies with the
best measured masses and dispersions (our Galaxy, M32, N4261, N4486,
N4564, N4697, and N7052), and find the same fit within the errors. The
68\% confidence limits for the best-fit linear relation come from
standard least-squares fitting. The uncertainty, based on Monte Carlo
simulations, is $\pm0.3$ in the exponent and 0.057~dex in the
normalization at 200~\kms.


\vskip 0.3cm

\begin{figure*}[t]
\centerline{\psfig{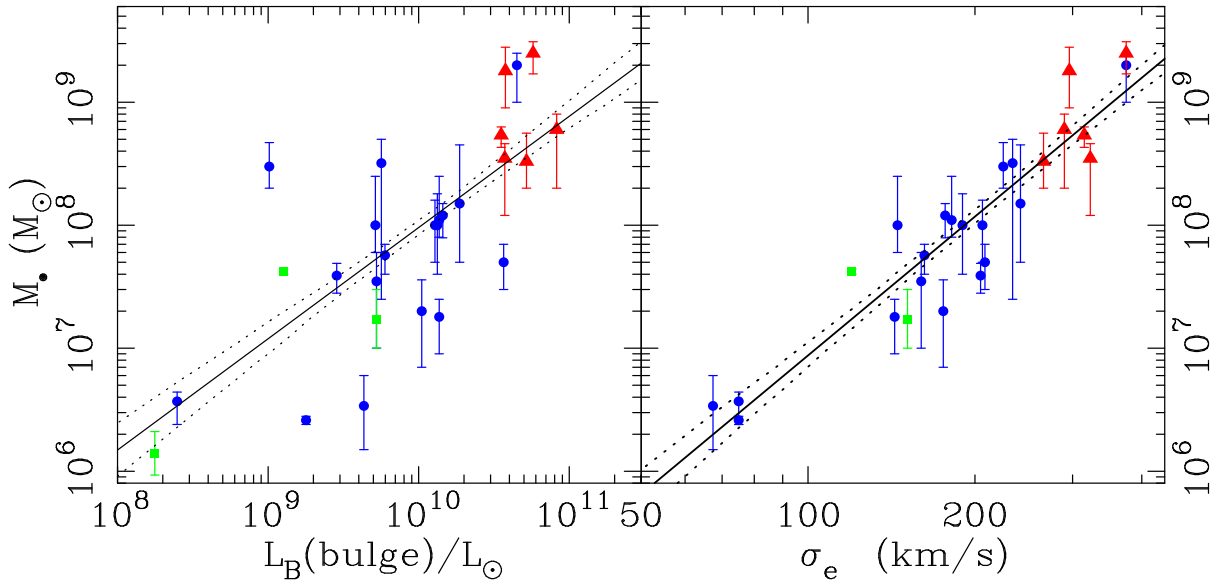}}
\figcaption[gebhardt.fig2.ps]{ Black hole mass versus bulge luminosity
(left panel) and the luminosity-weighted aperture dispersion within
the effective radius (right panel). There are 26 points in the
dispersion plot; 13 are new detections from stellar kinematics
(Gebhardt~\etal\ 2000b, Bower~\etal\ 2000). Green squares denote
galaxies with maser detections, red triangles come from gas
kinematics, and blue circles are from stellar kinematics. Solid and
dotted lines are the best-fit correlations and their 68\% confidence
bands.
\label{fig2}}
\end{figure*}


The measured scatter is 0.30 dex in black hole mass at fixed
dispersion. For the scatter estimate we use the biweight (Beers~\etal\
1990), and for this sample the biweight value is close to that of the
standard deviation. Given the likely measurement errors, the intrinsic
scatter is probably $<$ 0.15 dex. In fact, the scatter from the seven
galaxies with the best-measured quantities is 0.14 dex in black hole
mass. More accurate black hole masses are required to better constrain
the intrinsic scatter. With present data the intrinsic scatter is
consistent with zero.

It is possible that the tight correlation between black hole mass and
dispersion arises in part because the sample that we analyze consists
mainly of galaxies in {\it HST} observing programs that have been
chosen to look ``normal'' (e.g., no unusual morphology, active nuclei,
disturbed gas kinematics). It will be important to test the
correlation using a broader sample.

The galaxies that have the most leverage on the slope of the best-fit
correlation are those at the extremes, both low and high mass. We
discuss each of these, but we note that, when the three smallest and
two largest black holes are excluded, the best-fit slope still lies
within the original error bars. The three galaxies with the smallest
black hole masses are the Milky Way, M32, and NGC~7457. The masses for
our Galaxy (Ghez~\etal\ 1998; Genzel~\etal\ 2000) and M32
(van~der~Marel~\etal\ 1998) are among the most accurate of any in the
sample. The mass for NGC~7457 (discussed in Gebhardt~\etal\ 2000b) is
difficult to measure because there is a point source in the nucleus
(Lauer~\etal\ 1991); it may be overestimated. Measuring the
line-of-sight dispersions for NGC~7457 and M32 is straightforward
(Pinkney~\etal\ 2000b; van~der~Marel~\etal\ 1998) since in both
galaxies the dispersions are nearly independent of radius.  The
situation for our Galaxy is more difficult since we have to use the
near-infrared surface-brightness profile (Kent~\etal\ 1991) and
individual stellar velocities (Kent 1991; Genzel~\etal\ 2000) to
determine the aperture dispersion. The effective radius of the bulge
is uncertain due to disk contamination, but choosing values between 50
and 500 arcseconds provides dispersions between 85 and 65~\kms. We
adopt 75~\kms.

The two galaxies with the largest black holes are M87 (NGC~4486) and
NGC~4649. The black hole mass in M87 appears well determined.  The
mass of the black hole in NGC~4649 has substantial uncertainties since
its nuclear velocity dispersion is so large that absorption lines in
the STIS spectrum are washed out (Gebhardt~\etal\ 2000b). The
effective velocity dispersions in both of these galaxies have been
measured by several groups with good agreement.

We have looked for ways to reduce the scatter in the relation still
further, but without much success. Including the effective radius as
an additional parameter does provide a small improvement but only if
the dependence is very weak, $M_{\bullet} \propto R_e^{-0.1}$ at fixed
$\sigma_e$. We also examined the correlation of \mbh\ with aperture
dispersions within various radii, and find a similar correlation but
with slightly larger scatter---by 0.03 dex.

\section{Discussion}

A preliminary version of these results was presented at the June 2000
American Astronomical Society meeting in Rochester (Kormendy~\etal\
2000), where we learned of a similar analysis by Ferrarese \& Merritt
(2000). Their result, based on 12 objects, has a steeper slope
($M_{\bullet} \propto \sigma^{5.2}$).  This difference may be due to
different dispersions, black hole masses, or distances for the same
objects.  Our sample size is larger (26 compared to 12 galaxies), and
many of our black hole masses come from as-yet-unpublished
high-quality {\it HST} data and analyses. For their nearest galaxies
(the Galaxy and M32) it is likely that the central dispersions used by
Ferrarese \& Merritt (2000) are enhanced by the black hole, which in
turn creates a steeper slope. If we restrict our sample to the 11
galaxies common to their sample and ours but use our black hole masses
and dispersions, we find the same slope as presented here
($M_{\bullet}\propto \sigma_e^{3.75}$). Thus, we are confident that
our result is robust.

The implications of our results can be discussed in the context of the
fundamental plane for galaxies.  Roughly speaking, elliptical galaxies
and bulges can be described by three parameters: the effective radius
$R_e$, the total luminosity $L$, and the luminosity-weighted velocity
dispersion $\sigma$. If these galaxies have similar luminosity and
mass distributions and if the mass-to-light ratio is a well behaved
function of $R_e$ and $L$, then the virial theorem implies that they
occupy a two-dimensional manifold in the three-dimensional space with
coordinates $(\log L,\log\sigma,\log R_e)$; this is the so-called
``fundamental plane'' (Dressler et al. 1987; Djorgovski \& Davis 1987;
Faber et al. 1987). Any two of these variables can be used to predict
the value of the third.

Now let us examine the location of elliptical galaxies and bulges in
the four-dimensional space with coordinates $(\log M_{\bullet},\log
L,\log\sigma,\log R_e)$. The correlation of \mbh\ with $\sigma$ that
we describe here implies that (i) galaxies are still restricted to a
two-dimensional manifold---a fundamental plane---in this
four-dimensional space, and (ii) when projected onto the $(\log
M_{\bullet},\log\sigma)$ plane, the fundamental plane is viewed nearly
edge-on---in other words, contours of constant \mbh\ on the
fundamental plane are parallel to contours of constant $\sigma$. It is
remarkable that the scatter normal to the plane, 0.05 dex in three
dimensions, is increased to only 0.08 dex in four dimensions.

Most earlier discussions of the demography of central black holes
focused on the correlation of black hole mass with galaxy luminosity,
which can be regarded as a projection of the four-dimensional
fundamental plane onto the $(\log M_{\bullet},\log L)$ plane. We now
understand that most of the substantial scatter in that relation---0.6
dex in $\log M_{\bullet}$---is not a reflection of stochastic
processes that controlled the growth of black holes but arises instead
simply because in this projection the fundamental plane is not viewed
edge-on. Confirmation of this point is provided by the observation
that those galaxies which are outliers in the $(M_{\bullet},L)$
relation are not outliers in the $(M_{\bullet},\sigma_e)$ relation.

The tight correlation between black hole mass and velocity dispersion
strongly suggests a causal connection between the formation and
evolution of the black hole and the bulge. However, the nature of this
connection remains obscure (Haehnelt~\& Rees 1993; Haiman \& Loeb
1998; Silk~\& Rees 1998; Kauffman \& Haehnelt 2000).  It is natural to
assume that bulges, black holes, and quasars formed, grew, or turned
on as parts of the same process, in part because the collapse or
merger of bulges might provide a rich fuel supply to a centrally
located black hole. The finding that black hole mass correlates more
closely with dispersion than total luminosity or other global
properties probably reflects the fact that the aperture dispersion is
less sensitive to the properties of the outer galaxy.

We have shown that black hole mass is tightly coupled to velocity
dispersion of the host galaxy over three orders of magnitude. It
remains to be seen whether this correlation applies at still larger or
smaller black hole masses. It should be straightforward to explore
smaller masses, but examining higher-dispersion galaxies (e.g., cD
galaxies) will be challenging because of their low surface
brightnesses and large dispersions. We may even speculate whether the
correlation extends to stellar systems with dispersions as low as
dwarf spheroidal galaxies and globular clusters, although in other
respects these objects are quite different from bulges and elliptical
galaxies (Kormendy 1985). A typical dwarf spheroidal galaxy or massive
globular cluster has a dispersion of about 10~\kms; if the correlation
applies, it suggests a black hole mass of about
$2\times10^3~\Msun$. It is intriguing that Gebhardt~\etal\ (2000c)
measure an increase in the mass-to-light ratio of M15 near its center,
consistent with a central mass of 2500~$\Msun$.

\acknowledgements

We thank Avi Loeb for suggesting that we examine the correlation
between black hole mass and velocity dispersion. This work was
supported by {\it HST} grants to the Nukers, GO--02600.01--87A,
G06099, and G07388, and by NASA grant G5-8232. A.V.F. acknowledges
NASA grant NAG5-3556. K.G. is supported by NASA through Hubble
Fellowship grant HF-01090.01-97A awarded by the Space Telescope
Science Institute, which is operated by the Association of the
Universities for Research in Astronomy, Inc., for NASA under contract
NAS 5-26555.

\end{document}